\documentclass[onecolumn,amsmath,amssymb,aps,floatfix,showkeys,showpacs,12pt
]{revtex4-1}

\usepackage[T1]{fontenc}
\usepackage[utf8]{inputenc}

\usepackage{graphicx,colordvi,longtable,color}
\usepackage{dcolumn}
\usepackage{bm}

\begin{document}

\title{Single soliton solution to the extended KdV equation over uneven depth
}

\author{George Rowlands} \email{G.Rowlands@warwick.ac.uk}
\affiliation{Department of Physics, University of Warwick, Coventry, CV4 7A, UK}

\author{Piotr Rozmej}
 \email{P.Rozmej@if.uz.zgora.pl}
\affiliation{Institute of Physics, Faculty of Physics and Astronomy \\
University of Zielona G\'ora, Szafrana 4a, 65-246 Zielona G\'ora, Poland}

\author{Eryk Infeld} \email{Eryk.Infeld@ncbj.gov.pl}
\affiliation{National Centre for Nuclear Research, Hoża 69, 00-681 Warszawa, Poland}

\author{Anna Karczewska}
 \email{A.Karczewska@wmie.uz.zgora.pl}
\affiliation{Faculty of Mathematics, Computer Science and Econometrics\\ University of Zielona G\'ora, Szafrana 4a, 65-246 Zielona G\'ora, Poland}

\date{\today}

\begin{abstract}
In this note we look at the influence of a shallow, uneven riverbed on a soliton. The idea consists in approximate transformation of the equation governing wave motion over uneven bottom to equation for flat one for which the exact solution exists.
The calculation is one space dimensional, and so corresponding to long trenches or banks under wide rivers or else oceans.
\end{abstract}



\maketitle


\section{Introduction}
Recently, we found exact solitonic \cite{KRI14} and periodic  \cite{IKRR17} wave solutions for water waves moving over a smooth riverbed. Amazingly they were simple, though governed by a more exact expansion of the Euler equations with several new terms as compared to KdV 
\cite{MS90,BS13,KRR14,KRI14}. Our next step is to consider how these results are modified by a rough river or ocean bottom. We start with a simple case. The geometry is one space dimensional and the wave a soliton. Even so, approximations rear their 
head! Considerations of a two dimensional bump on the bottom, as well as periodic waves propagating overhead, are planned for a later effort.

Here we consider the following equation governing the elevation of the
water surface $\eta/H$ above a flat equlibrium at the surface (written in dimensionless variables)
\begin{align} \label{kdv2d}   \eta_t+\eta_x + \alpha \frac{3}{2}\eta\eta_x +\beta\frac{1}{6} \eta_{3x} + \alpha^2\left(\!-\frac{3}{8}\eta^2\eta_x \!\right)  + \alpha\beta\left(\!\frac{23}{24}\eta_x\eta_{2x}\!+\!\frac{5}{12}\eta\eta_{3x}\! \right)+\beta^2\frac{19}{360}\eta_{5x} && \\  +\beta\delta \frac{1}{4}\left(\!-\frac{2}{\beta}(h\eta)_x \!+\! (h_{2x}\eta)_x \!-\! (h\eta_{2x})_x\! \right) &=&0. \nonumber \end{align}
The last three terms are due to a bottom profile.
We emphasize, that (\ref{kdv2d}) was derived in \cite{KRR14,KRI14} under the  assumption that $\alpha,\beta,\delta$ are small (positive by definition) and of the same order.
As usual, $\alpha=A/H$, the ratio of wave amplitude $A$ to mean water depth $H$ and $\beta= (H/L)^2$ where $L$ is mean wavelength. 
Parameter $\delta=A_h/H$ is the ratio of the amplitude of the bottom function $h(x)$ to mean water depth. Up to this point $A,H,L, A_h$ are dimension quantities. Scaling to dimensionless variables allows us to apply perturbation approach to the set of Euler equations governing the model of ideal fluid. In the first order perturbation approach the KdV equation is obtained (assuming flat bottom). Applying second order perturbation approach Marchant and Smyth \cite{MS90} derived equation (\ref{kdv2d}) limited to the first line and named the \emph{extended KdV equation}. Since it is derived in second order perturbation with respect to small parameters we call it KdV2. Taking into account small bottom fluctuations (again in second order perturbation approach)
led us in \cite{KRR14,KRI14} to the KdV2 equation for uneven bottom (\ref{kdv2d}).
In scaled variables amplitudes of wave and bottom profiles are equal one. 
In \cite{KRI14,IKRR17} we derived exact soliton and periodic solutions to KdV2. These solutions are given by the same functions as the corresponding KdV solutions but with different coefficients. 

This paper presents an attempt to describe dynamics of the exact KdV2 soliton when it approches a finite interval of uneven bottom. 
We will use reductive perturbation method introduced by Taniuti and Wei \cite{TW68}. Using two space scales allows us to transform equation for uneven bottom (\ref{kdv2d}) into KdV2 equation with some coefficients altered, that is, equation for the flat bottom. This transformation is approximate but analytical solution of the resulted equation is known. This approximate analytic description will be compared with exact numerical calculations.

\section{KdV2 soliton (even bottom)}

In this section we shortly remind exact soliton solution of the KdV2 equation given in~\cite{KRI14}.

Assume the form of a soliton moving to the right, $\eta(x,t)=\eta(x-vt)$. So, $\eta_t=-v\,\eta_x$ and 
 the KdV2 equation, that is (\ref{kdv2d}) without the last row, becomes ODE 
\begin{equation} \label{eta1}
(1-v)\eta_x + \alpha\, \frac{3}{2}\eta\eta_x +\beta\,\frac{1}{6} \eta_{3x} -\frac{3}{8} \alpha^2 \eta^2\eta_x +
 \alpha\beta\,\left(\frac{23}{24}\eta_x\eta_{2x}+\frac{5}{12}\eta\eta_{3x} \right)+\beta^2\,\frac{19}{360}\eta_{5x} =0. \quad
\end{equation} 
Integration gives
\begin{equation} \label{eta2}
(1-v)\eta + \alpha\, \frac{3}{4}\eta^2 +\beta\,\frac{1}{6} \eta_{2x} -\frac{1}{8}\alpha^2 \eta^3 +
 \alpha\beta\,\left(\frac{13}{48}\eta_x^2+\frac{5}{12}\eta\eta_{2x} \right)+\beta^2\,\frac{19}{360}\eta_{4x} =0. 
\end{equation}
Then the solution is assumed in the same form as KdV solution 
\begin{equation} \label{1sol}
\eta(y)= A\,\text{Sech}^2(By),
\end{equation}
where $A=1$, since in dimensionless variables the amplitude is already rescalled. However, for further consideretions it is convenient to keep the general notation. Insertion postulated form of the solution  (\ref{1sol}) and use of properties of hyperbolic functions gives (\ref{eta2}) in polynomial form
\begin{equation} \label{war1}
C2\, \mbox{Sech}^2(By) + C4\, \mbox{Sech}^4(By) +C6\, \mbox{Sech}^6(By)=0,
\end{equation}
which requires simultanoeus vanishing of all coefficients $C2,C4,C6$. These three conditions are as follows
\begin{eqnarray} \label{c2}
(1-v) + \frac{2}{3}  B^2 \beta + \frac{38}{45}  B^4 \beta^2 &=& 0, \\ \label{c4}
\frac{3  A\alpha}{4} -  B^2 \beta + \frac{11}{4} A\alpha\, B^2  \beta -  \frac{19}{3} B^4 \beta^2 &=& 0,\\ \label{c6}
-\left(\frac{1}{8}\right) (A\alpha)^2 - \frac{43}{12} A\alpha\, B^2  \beta +  \frac{19}{3}  B^4 \beta^2 &=& 0.
\end{eqnarray}
Denoting $\displaystyle z=\frac{\beta B^2}{\alpha A}$ one obtains (\ref{c6}) as quadratic equation with respect to $z$ with solutions
\begin{equation} \label{rkw1}
 z_1 = \frac{43-\sqrt{2305}}{152}\approx -0.033<0 \qquad \mbox{and} \qquad
 z_2  =  \displaystyle \frac{43+\sqrt{2305}}{152}\approx 0.599 > 0. 
\end{equation}
Since $B=\sqrt{\frac{\alpha}{\beta}zA}$, only $z_2$ provides real $B$ value.  [In principle $\text{Sech}^2$ of imaginary argument can be expressed by a quotient of expressions given by hyperbolic functions of real arguments. However, these expressions are singular for some values of arguments and therefore physically irrelevant.]
 
 Eqs. (\ref{c4}) and (\ref{c6}) are consistent only when
$\displaystyle \alpha=\alpha_s=\frac{3(51-\sqrt{2305})}{37}\approx 0.242399$.
Then (\ref{c2}) determines velocity
\begin{equation} \label{vv1}
v=1+\frac{2}{3} \alpha_s z_2+\frac{38}{45} (\alpha_s z_2)^2 \approx 1.114546.
\end{equation}

\section{Variable depth}\label{RB}

 Equation  (\ref{kdv2d}) can be written in the form
\begin{equation} \label{basic}
\frac{\partial \eta}{\partial t} +\frac{\partial}{\partial x} f(\eta,h) = 0,
\end{equation}
 where $f(\eta,h)$ is given by 
\begin{align} \label{bas0} f(\eta,h) = ~& \eta +\frac{3\alpha}{4} \eta^2 -\frac{\alpha^2}{8} \eta^3+\alpha\beta\left[\frac{13}{48} \left(\frac{\partial \eta}{\partial x} \right)^2 +\frac{5}{12} \eta \frac{\partial^2 \eta}{\partial x^2} \right] + \frac{\beta}{6} \frac{\partial^2 \eta}{\partial x^2}  + \frac{19}{360} \beta^2 \frac{\partial^4 \eta}{\partial x^4} \nonumber \\ & +\beta\delta\left[-\frac{2}{\beta}h\eta + \frac{\partial^2 h}{\partial x^2} \eta- h\frac{\partial^2 \eta}{\partial y^2} \right] .   \end{align}  
We treat $h$ as slowly varying and introduce two space scales $x$ and $x_1(=\epsilon x)$ which are treated as independent until the end of calculation \cite{TW68}
\begin{equation} \label{hx1}
 h = h(\epsilon x) = h(x_1) \qquad \epsilon \ll 1.
\end{equation}
We also introduce
\begin{equation} \label{yy}
y=\int_0^x a(\epsilon x) dx   - t
\end{equation}
where $a$ is as yet undefined. To first order in $\epsilon$
\begin{equation} \label{eta} 
\eta=\eta_0(y,x_1) +\epsilon\eta_1(y,x_1) + \ldots
\end{equation}
\begin{equation} \label{etat} 
\frac{\partial \eta}{\partial t} = - \frac{\partial \eta_0}{\partial y} 
- \epsilon\frac{\partial \eta_1}{\partial y} + \ldots \end{equation}

\begin{equation} \label{etax} 
\frac{\partial \eta}{\partial x} = a(x_1) \frac{\partial \eta_0}{\partial y}  + \epsilon\frac{\partial \eta_0}{\partial x_1} + \epsilon a(x_1) \frac{\partial \eta_1}{\partial y} + \ldots \end{equation}
\begin{equation} \label{eta2x}
\frac{\partial^2 \eta}{\partial x^2} = a^2 \frac{\partial^2 \eta_0}{\partial y^2} + \epsilon\left( \frac{\partial a}{\partial x_1}\frac{\partial \eta_0}{\partial y} +2a\frac{\partial^2 \eta_0}{\partial y \partial x_1}+a^2 \frac{\partial^2 \eta_0}{\partial y^2}\right)
 + \ldots \end{equation}

Now 
\begin{equation}  \label{etanx}
\frac{\partial^n \eta}{\partial x^n} = a^n \frac{\partial^n \eta_0}{\partial y^n} + O(\epsilon)\,.  \end{equation}
We have 
\begin{align} \label{bas1} f(\eta,h) &= \eta_0 +\frac{3\alpha}{4} \eta_0^2 -\frac{\alpha^2}{8} \eta_0^3+\alpha\beta\left[\frac{13}{48} a^2 \left(\frac{\partial \eta_0}{\partial y} \right)^2 +\frac{5}{12} \eta_0 a^2 \frac{\partial^2 \eta_0}{\partial y^2} \right] + \frac{\beta}{6} a^2 \frac{\partial^2 \eta_0}{\partial y^2}  + \frac{19}{360} \beta^2 a^4 \frac{\partial^4 \eta_0}{\partial y^4} \nonumber \\ & +\beta\delta\left[-\frac{h(x_1)\eta_0}{2\beta}- a^2 h(x_1)  \frac{\partial^2 \eta_0}{\partial y^2} \right] +O(\epsilon) = f_0(\eta_0,h) +O(\epsilon) \,. \end{align}
 
 From (\ref{basic}), (\ref{etat}) and (\ref{bas1}) to lowest order we have
\begin{equation} \label{bas2} 
-\frac{\partial \eta_0}{\partial y} + a \frac{\partial }{\partial y}\left[f_0(\eta_0,h) \right] =0 \end{equation}
and, since $a=a(x_1)$,  we obtain 
\begin{equation} \label{de0y}
\frac{\partial }{\partial y}(\eta_0-a f_0 ) =0 \,.
\end{equation}

We restrict consideration to a single soliton, so $\eta_0\to 0$ as $y\to \pm\infty$ and so does $f_0$. Integration of (\ref{de0y}) yields to lowest order 
\begin{equation} \label{e0f0}
\eta_0-  a(x_1) f_0 = 0 \,.
\end{equation}

Introduce $\zeta=y/a(x_1)$  which is constant in our approximation. Now
 \begin{equation} \label{de0z}
\frac{\partial \eta_0}{\partial y} = \frac{1}{a}  \frac{\partial \eta_0}{\partial \zeta} \end{equation}  
and from (\ref{e0f0}), (\ref{bas1}), (\ref{de0z}) we obtain
\begin{eqnarray} \label{feh1}
 (1-a(x_1))\eta_0 - \frac{3\alpha}{4} \eta_0^2 \,a +\frac{\alpha^2}{8} \eta_0^3  \,a- \alpha\beta\left[\frac{13}{48}  \left(\frac{\partial \eta_0}{\partial \zeta}\right)^2 +\frac{5}{12} \eta_0 \frac{\partial^2 \eta_0}{\partial \zeta^2}\right] \,a  && \nonumber \\
 -\frac{19}{360} \beta^2 \frac{\partial^4 \eta_0}{\partial \zeta^4} \,a
 +\beta\delta\, h(x_1)\left[\frac{\eta_0}{2\beta}+\frac{\partial^2 \eta_0}{\partial \zeta^2} \right] \,a -  \frac{\beta}{6} \frac{\partial^2 \eta_0}{\partial \zeta^2} \,a &=& 0 \,.
\end{eqnarray}
 Dividing by $(-a)$ yields 
\begin{eqnarray} \label{feh2}
\left(1- \frac{\delta\, h}{2}-\frac{1}{a}\right)\eta_0 +\frac{3\alpha}{4} \eta_0^2 - \frac{\alpha^2}{8} \eta_0^3 +  \alpha\beta\left[\frac{13}{48}  \left(\frac{\partial \eta_0}{\partial \zeta}\right)^2 + \frac{5}{12} \eta_0  \frac{\partial^2 \eta_0}{\partial \zeta^2}\right] &&   \\
+ \frac{19}{360} \beta^2  \frac{\partial^4 \eta_0}{\partial \zeta^4} + \frac{\beta}{6}\left(1- 
6\,\delta\,h\right)\frac{\partial^2 \eta_0}{\partial \zeta^2} &=& 0. \nonumber
\end{eqnarray}
This should be compared to (\ref{eta2}) or  \cite[Eq.~(22)]{KRI14}. Remember that at this stage $\delta\, h(x_1)$ is to be treated as constant with respect to inegration over $\zeta$. The only difference is that ~$\displaystyle v=\left(\frac{\delta h}{2} +\frac{1}{a}\right)$~ and  ~$(1-
6 \,\delta \,h)$~  instead of 1 appear in  the last term. 

Following  \cite{KRI14} we obtain 
\begin{equation} \eta_0 = A\, \text{sech}^2(B\,\zeta),  \qquad \zeta=\frac{1}{a(x_1)}\left[\int_0^x a(x_1) dx-t\right]. \end{equation}
In equations  (\ref{c2})-(\ref{c4}) [(24), (25)  and (20) of \cite{KRI14}] 
we replace  $\beta\,B^2$  (but not $\beta^2 B^4$ or $\alpha \beta A B^2$ since we modify only first order terms) by 
\begin{equation}\displaystyle \beta\left( 1-
6\,\delta\,h\right)B^2.\end{equation}
Now $z=z_2=\frac{43+\sqrt{2305}}{152}$ is as in (\ref{rkw1}). 
We obtain
\begin{equation} \label{apps}
\eta_0 = \bar{A} \,\text{sech}^2\left[\frac{\bar{B}}{a(x_1)}\left(\int_0^x a(x_1)dx-t\right) \right]
\end{equation} 
with 
\begin{equation} \frac{1}{a} +\frac{\delta \,h}{2} = v -\beta \delta h , \qquad q=\frac{b}{B^2}, \qquad b=\frac{3z}{\frac{76}{3}z-11}
\end{equation}

and 
$$ \bar{A}=A(1+q\delta h), \qquad \bar{B}=B(1+q\delta\, h/2),
$$
where $A,B,v$ are given by eqs. (30)-(32) in \cite{KRI14}. 
Thus
\begin{equation}  \frac{1}{a} = v-\left(\frac{1}{2}+\beta \right) \delta h .\end{equation}
At this stage we take  $x_1=\epsilon \,x$~ and ~$ \delta h= \delta h(x)$.
So
\begin{equation} \int_0^x a(x)\, dx = \int_0^x \frac{dx}{v-\left(\frac{1}{2}+\beta \right) \delta h(x)}. \end{equation}

Assume $\delta h(x)$ is nonzero only in interval  ~$x\in[L_1,L_2]$.

For $x <L_1, \quad \eta_0=A \text{sech}^2(B(x-vt)), \quad \delta h\equiv 0,~ \frac{1}{a}=v$ .

For $x >L_2, \quad \delta h\equiv 0,~ \frac{1}{a}=v$~ and 
\begin{equation} \eta_0=A\, \text{sech}^2\left[ B \left(v\int_{L_1}^{L_2} a(x)dx +(x-vt) \right)\right]\,.  \end{equation}

There is a change of phase as the pulse passes through the region where $\delta h\ne 0$. The alteration in the phase is given by 
\begin{equation} \int_{L_1}^{L_2} dx\left[\frac{1}{1-\frac{(1/2+\beta)\delta h}{v}}-1 \right]
\approx \frac{\beta+1/2}{v}\int_{L_1}^{L_2} \delta \,h(x)\,dx \,.
\end{equation}
If this integral is zero  phase is unaltered. This can happen if a deeper region is followed by a shallower region of appropriate shape or vice versa.

\subsection{Examples}

In the following figures we present time evolution of the approximate analytic solution (\ref{apps}) to KdV2 equation with uneven bottom (\ref{kdv2d}) for several values of parameters of the system. These evolutions are compared with 'exact' numerical solutions of (\ref{kdv2d}). In both cases initial conditions were the exact solutions of KdV2 equation. Therefore in all presented examples  $\alpha=\alpha_s$ and the amplitude of initial soliton is equal to 1.

\begin{figure}[tbh]
\begin{center}
\resizebox{0.8\columnwidth}{!}{\includegraphics{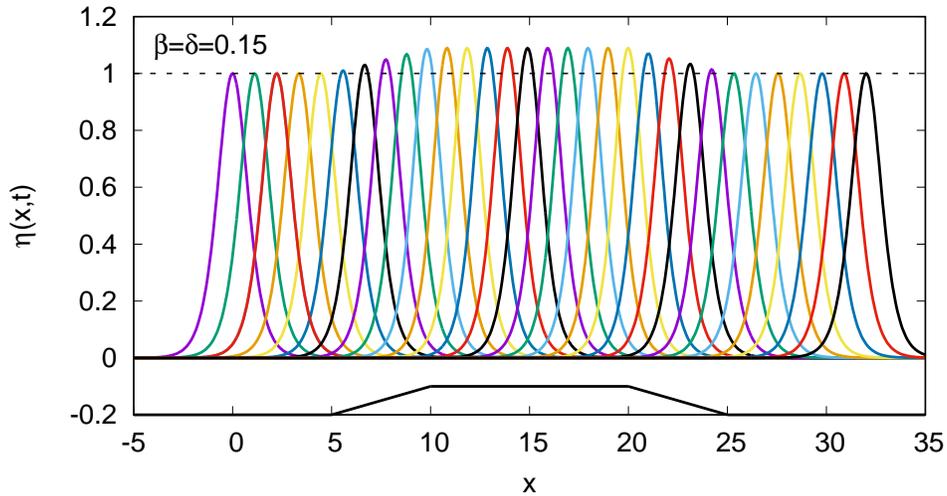}}
\end{center}
\vspace{-8mm}
\caption{Profiles of the soliton as given by (\ref{apps}). The shape of the trapezoidal bottom is shown (not in scale). Consecutive times are $t_n=n$,  ~$n=0,1,2,3,\ldots,32$.} \label{bd015A}  
\end{figure}
\vspace{-5mm}
\begin{figure}[bht]
\begin{center}
\resizebox{0.8\columnwidth}{!}{\includegraphics{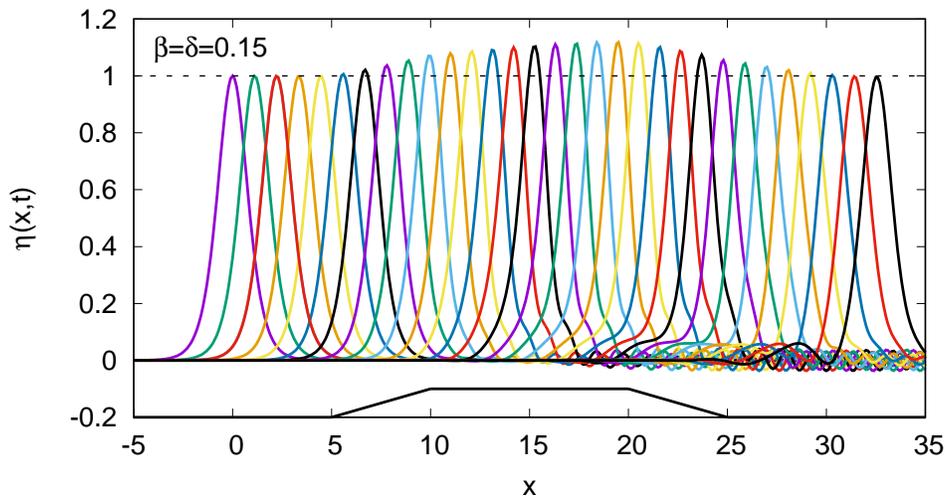}}
\end{center}
\vspace{-6mm}
\caption{Profiles of numerical solution of the equation (\ref{kdv2d}) obtained with the same initial condition. Time instants the same as in fig.~\ref{bd015A}. } \label{bd015N}  
\end{figure}

In figure \ref{bd015A} we present the approximate solution (\ref{apps}) for the case when soliton moves over a trapezoidal elevation with $L_1=5$ and $L_2=25$. We took  ~$\beta=\delta=0.15$. For smaller $\delta$ the effects of uneven bottom are very small, for larger $\delta$ second order effects (not present in analytic approximation) cause stronger overlaps of different profiles.

\begin{figure}[bht]
\begin{center}
\resizebox{0.8\columnwidth}{!}{\includegraphics{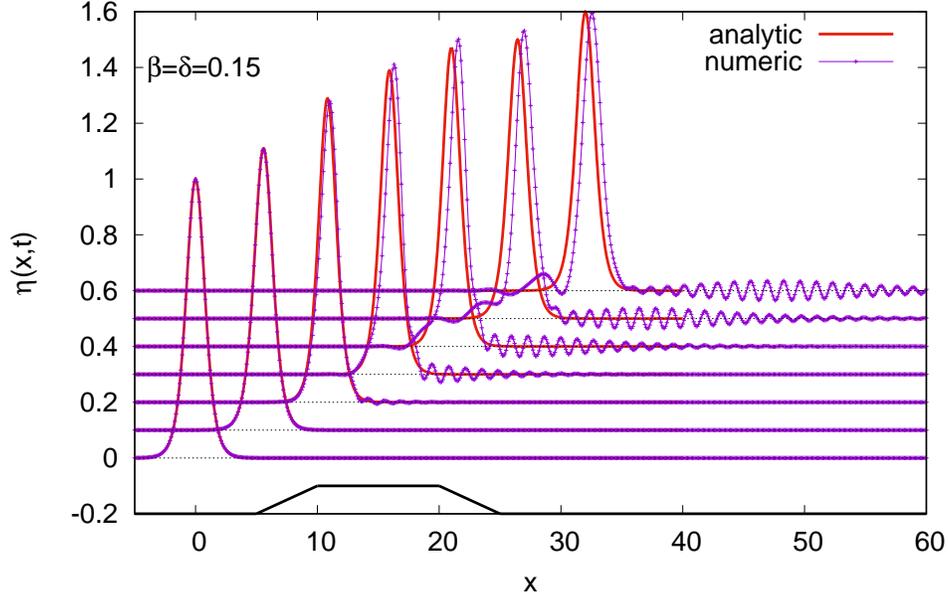}}
\end{center}
\vspace{-2mm}
\caption{Comparison of +48 68 3282 909wave profiles shown if figs.~\ref{bd015A} and \ref{bd015N} for  time instants $t=0,5,10,15,20,25,30$. Consecutive profiles are vertically shifted by 0.1} \label{bd015AN}  
\end{figure}

We compare this approximate solution of (\ref{kdv2d}) to a numerical simulation  obtained with the same initial condition. The evolution is shown in figure \ref{bd015N}. We see that the approximate solution has the main properties of the soliton motion as governed by equation (\ref{kdv2d}).
However, since the numerical solution contains higher order terms depending on the shape of $h$ the exact motion as obtained from numerics shows additional small amplitude structures known from earalier papers, for example \cite{KRR14,KRI14}. This is clearly seen in fig.~\ref{bd015AN} where profiles obtained in analytic and numeric calculations are compared at time instants
$t=0,5,10,15,20,25,30$ on wider interval of $x$. All numerical results were obtained with calculations performed on  wider interval $x\in [-30,70]$ with periodic boundary conditions.
Details of numerics was described in \cite{KRR14,KRI14,IKRR17}.

\vspace{-2mm}

\begin{figure}[tbh]
\begin{center}
\resizebox{0.8\columnwidth}{!}{\includegraphics{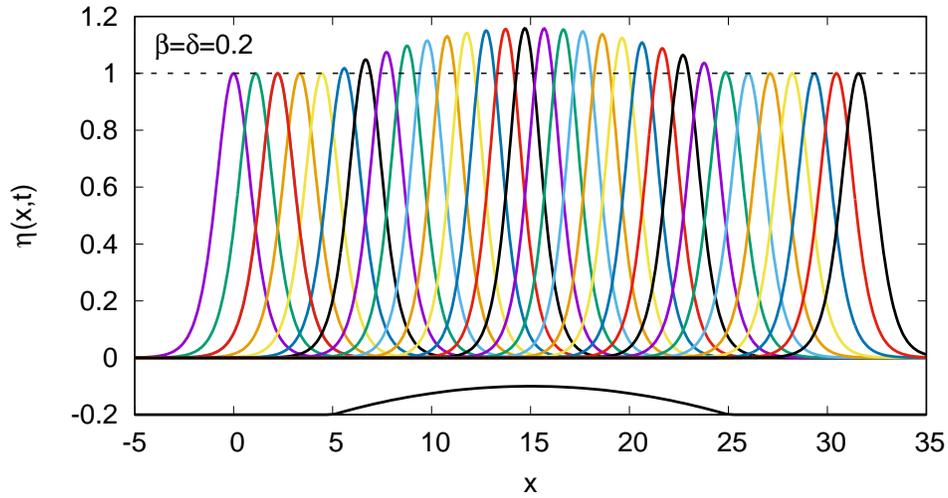}}
\end{center}
\vspace{-2mm}
\caption{Profiles of the soliton as given by (\ref{apps}). The shape of the trapezoidal bottom is shown (not in scale). Consecutive times are $t_n=n$,  ~$n=0,1,2,3,\ldots,32$.} \label{bd02AL}  
\end{figure}

\begin{figure}[bht]
\begin{center}
\resizebox{0.8\columnwidth}{!}{\includegraphics{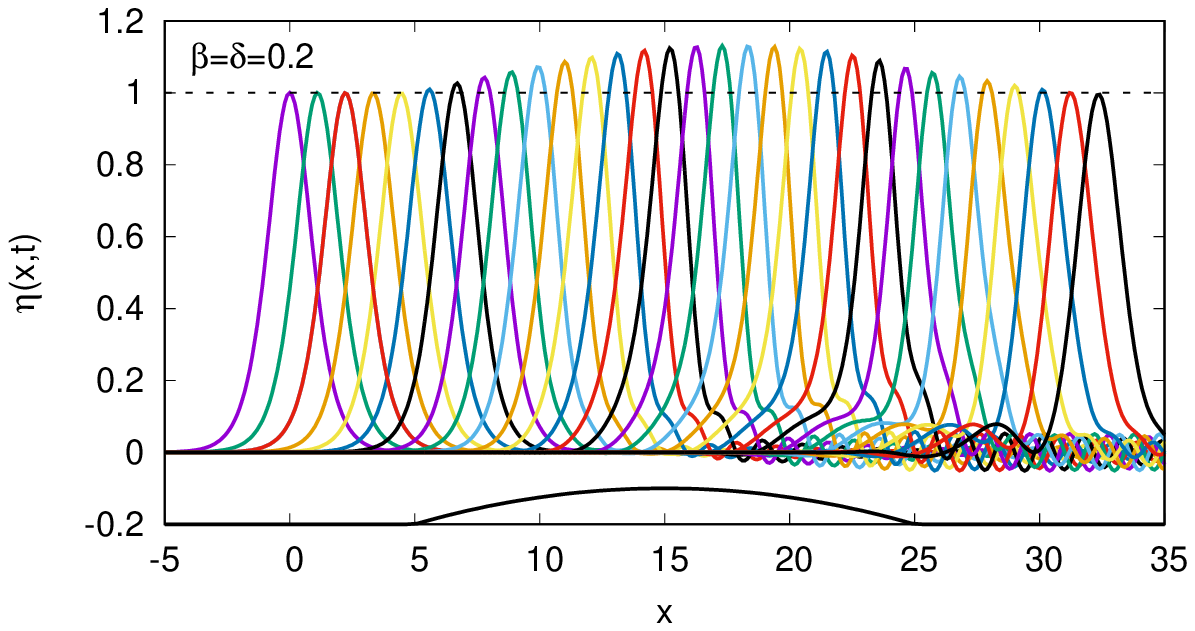}}
\end{center}
\vspace{-2mm}
\caption{Profiles of numerical solution of the equation (\ref{kdv2d}) obtained with the same initial condition. Time instants the same as in fig.~\ref{bd02AL}. } \label{bd02NL}  
\end{figure}

\begin{figure}[bht]
\begin{center}
\resizebox{0.8\columnwidth}{!}{\includegraphics{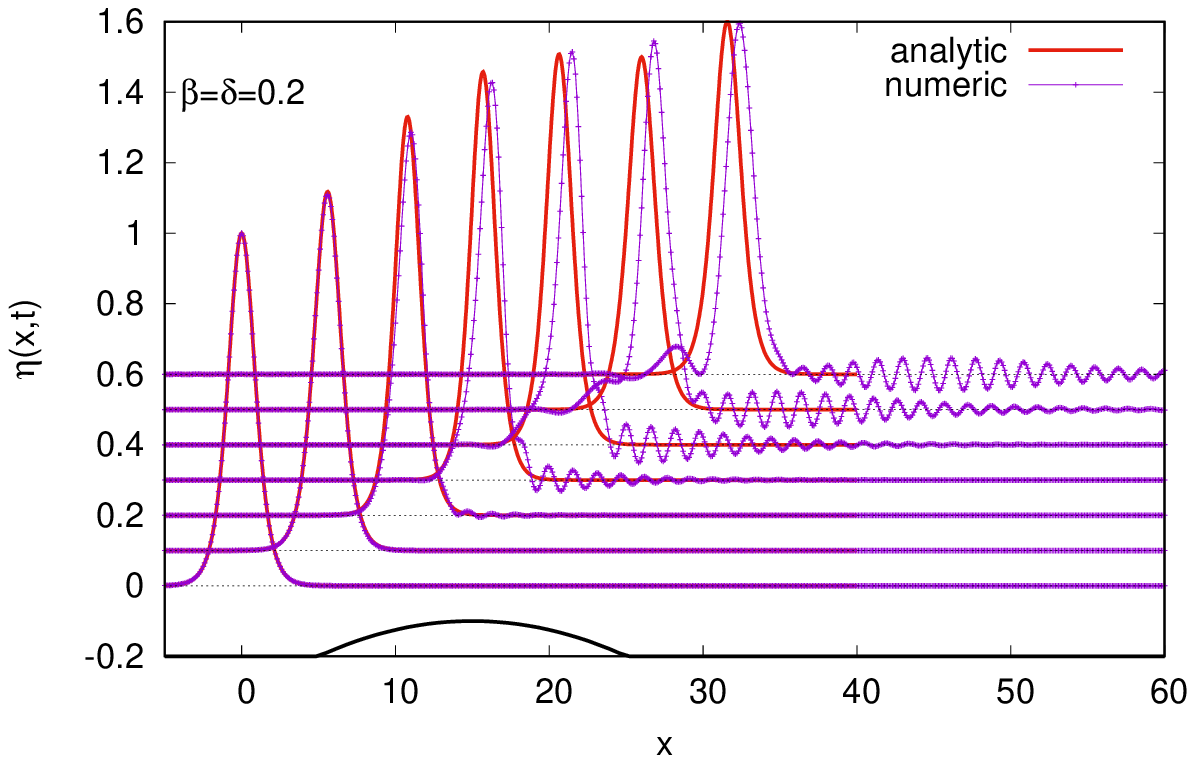}}
\end{center}
\vspace{-2mm}
\caption{Comparison of wave profiles shown if figs.~\ref{bd02AL} and \ref{bd02NL} for  time instants $t=0,5,10,15,20,25,30$. } \label{bd02ANLUK}  
\end{figure}

In figures \ref{bd02AL}-\ref{bd02ANLUK} we present results analogous to those presented in figures \ref{bd015A}-\ref{bd015AN} but with a different shape of the bottom bump and larger values of $\beta=\delta=0.2$. 
In this case the bump is chosen as an arc of parabola $h(x)=1-(x-15)^2/100$ between the same $L_1=5$ and $L_2=24$ as in trapezoidal case. 

\begin{figure}[thb]
\begin{center}
\resizebox{0.85\columnwidth}{!}{\includegraphics{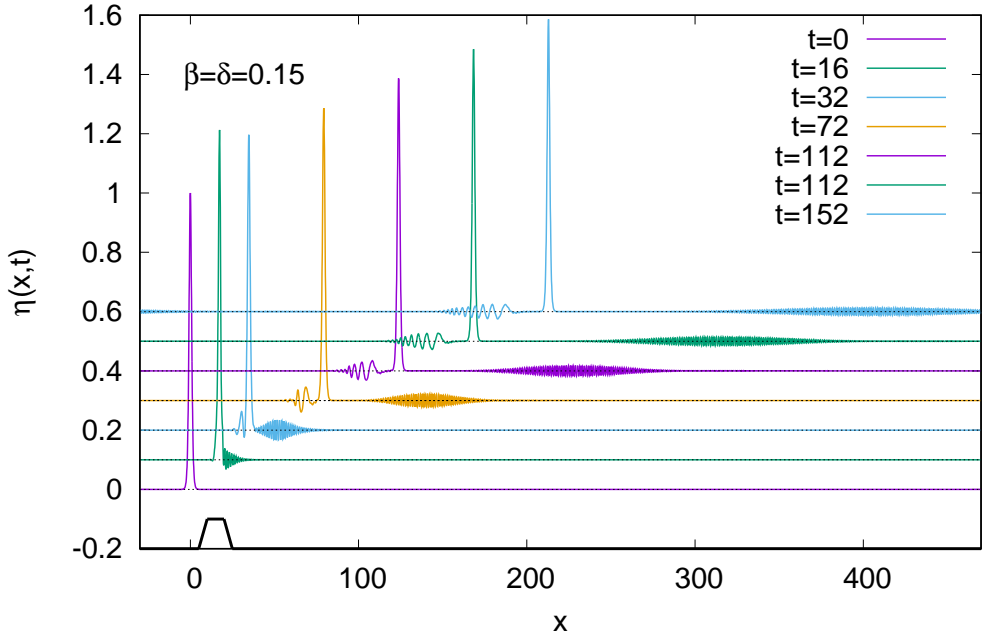}}
\end{center}
\vspace{-6mm}
\caption{Long time numerical evolution with trapezoidal bottom bump for $\beta=\delta=0.15$.} \label{f7}  
\end{figure}

In approximate analytic solution KdV2 soliton changes its amplitude and velocity only over bottom fluctuation. When the bottom bump is passed it comes back to initial shape (only phase may be changed). This is not the case for 'exact' numerical evolution of the same initial KdV2 soliton when it evolves according to the second order equation (\ref{kdv2d}). This is clearly visible in figures \ref{bd015AN} and \ref{bd02ANLUK}. What is this motion for much larger times? In order to answer this question one has to perform numerical calculations on much wider interval of $x$. Such results are presented in figure \ref{f7}. The interaction of soliton with the bottom bump creates two wave packets of small amplitudes. First moves with higher frequency faster than the soliton and is created when soliton enters the bump, second moves slower with lower frequency and appears when soliton leaves it. After some time both are separated from the main wave. Since periodic boundary conditions were used in numerical algorithm, the head of wave packet radiated forward travelled for $t=152$ larger distance than the interval chosen for calculation and is seen at left side of the wave profile.

We have to epmhasize that this behaviour is generic, it looks similar for different shapes of bottom bumps and different values of $\beta,\delta$ parameters.  It was observed in our earlier papers \cite{KRI14,KRI15,KRIR17} in which initial conditions were in the form of KdV soliton. 

\section{Conclusions}

 We have derived a simple formula describing approximately a soliton encountering an
uneven riverbed. The model reproduces the known increase in amplitude when 
passing over a shallower region,
as well as the change in phase. However, the full dynamics of the soliton motion is much richer, the uneven bottom causes low amplitude soliton radiation both ahead and after the main wave. This behaviour was observed in our earlier papers \cite{KRI14,KRI15,KRIR17} in which initial conditions were in the form of KdV soliton, whereas in the present cases the KdV2 soliton, that is, exact solution of the KdV2 equation was used. 

\vskip6pt


\end{document}